\documentclass[journal]{IEEEtran}
\usepackage{amsmath,amsfonts}
\usepackage{algorithmic}
\usepackage{algorithm}
\usepackage{array}
\usepackage[caption=false,font=normalsize,labelfont=sf,textfont=sf]{subfig}
\usepackage{textcomp}
\usepackage{stfloats}
\usepackage{url}
\usepackage{verbatim}
\usepackage{graphicx}
\usepackage{cite}
\usepackage{xcolor}
\usepackage{listings}
\usepackage{multirow} 

\definecolor{codegreen}{rgb}{0,0.6,0}
\definecolor{codegray}{rgb}{0.5,0.5,0.5}
\definecolor{codepurple}{rgb}{0.58,0,0.82}
\definecolor{backcolour}{rgb}{0.95,0.95,0.92}

\lstdefinestyle{mystyle}{
    backgroundcolor=\color{backcolour},   
    commentstyle=\color{codegreen},
    keywordstyle=\color{magenta},
    numberstyle=\tiny\color{codegray},
    stringstyle=\color{codepurple},
    basicstyle=\ttfamily\footnotesize,
    breakatwhitespace=false,         
    breaklines=true,                 
    captionpos=b,                    
    keepspaces=true,                 
    numbers=left,                    
    numbersep=5pt,                  
    showspaces=false,                
    showstringspaces=false,
    showtabs=false,                  
    tabsize=2
}
\begin{document}

\title{Talk with the Things: Integrating LLMs into IoT Networks}

\author{Alakesh Kalita,~\IEEEmembership{Senior Member,~IEEE,}

\thanks{} 
\noindent
\thanks{Alakesh Kalita is with the Department of Mathematics and Computing, Indian Institute of Technology (ISM) Dhanbad, India.}

\thanks{}
\thanks{}
}

\maketitle

\begin{abstract}
The convergence of Large Language Models (LLMs) and Internet of Things (IoT) networks opens new opportunities for building intelligent, responsive, and user-friendly systems. This work presents an edge-centric framework that integrates LLMs into IoT architectures to enable natural language-based control, context-aware decision-making, and enhanced automation. The proposed modular and lightweight Retrieval-Augmented Generation (RAG)-based LLMs are deployed on edge computing devices connected to IoT gateways, enabling local processing of user commands and sensor data for reduced latency, improved privacy, and enhanced inference quality.
We validate the framework through a smart home prototype using LLaMA 3 and Gemma 2B models for controlling smart devices. Experimental results highlight the trade-offs between model accuracy and inference time with respect to model size. 
At last, we also discuss the potential applications that can use LLM-based IoT systems, and a few key challenges associated with such systems. 
\end{abstract}


\section{Introduction}
\IEEEPARstart{W}\ ith the development of digitization, IoT (\textit{Internet of Things}) technology has become a prominent topic in the field of networking and communication. IoT aims to connect billions of physical devices, which can be both living and non-living entities found on the earth's surface, with sensing/actuation and communication capacities. 
By enabling seamless connectivity and data exchange, IoT is transforming how we interact with the world around us \cite{10.1145/3536166}. On the other hand, new deep generative-based \textit{Large Language Models} (LLMs) are designed to understand and generate human language. Recent advances in generative AI and transformer-based LLMs have demonstrated remarkable capabilities in understanding natural language and reasoning. LLMs, such as GPT-4, GPT-4o, and Llama 3.1, are trained on vast amounts of text data, enabling them to perform a wide range of language tasks, including translation, summarization, and conversation. 

Notably, traditional IoT networks only work commands and actions basis in the environment like smart home, industry etc. However, the integration of  LLMs in IoT networks can make a huge transformation towards more intelligent, efficient, and responsive IoT systems \cite{kok2024}, \cite{gao2024}, \cite{lin2024pushinglargelanguagemodels}. This convergence leverages the natural language understanding capabilities of LLMs in IoT, allowing devices to communicate in a more meaningful and context-aware manner \cite{an2025}. For example, if a user says ``\texttt{Set up for movie night}'', LLMs can help to \textit{set the Dim lights in the living room}, \textit{close the blinds, adjust the thermostat to a comfortable temperature, turn on the TV and navigate to the user's preferred streaming service}. All can be done with a single command, which requires multiple individual commands for each actuation in the current IoT systems. In brief,  IoT can serve as the ``\textit{sensing limbs}'' of an environment, feeding real-time sensor inputs (e.g., temperature, motion, vibrations) to LLMs, while LLMs act as the ``\textit{brain}'' that analyzes data, draws conclusions, and communicates or acts upon them. Thus, LLM-based communication in IoT can significantly improve efficiency, bandwidth utilization, and user interactions.

However, the devices used in IoT networks are resource-constrained in terms of processing capacity, memory, and power. Therefore, using LLM in this kind of device is not practically trivial. 
Similarly, technologies like cloud computing have their disadvantages, like higher latency and cost. Therefore, edge computing can be a suitable approach to use LLM in IoT applications such as smart homes, smart industries, etc. Edge computing processes data closer to the source of the data generators rather than relying on centralized data centres at higher propagation delay \cite{10342693}. Edge computing reduces latency, conserves bandwidth, and enhances real-time data processing capabilities. 

Therefore, in this article, we briefly discuss how LLMs can be used to establish efficient communication in edge computing-based IoT systems. In the next section, we present our proposed framework for integrating LLMs into edge-based IoT systems. We then describe our experimental setup in a smart home environment and discuss the results obtained. Finally, before concluding, we highlight the advantages of using LLMs in IoT systems, explore additional use cases beyond smart homes, and outline key challenges that must be addressed to make such systems more efficient and robust.

\section{System Architecture: How LLMs Work in an IoT Network}

\begin{figure*}[!t]
\centering
\includegraphics[width=\textwidth]{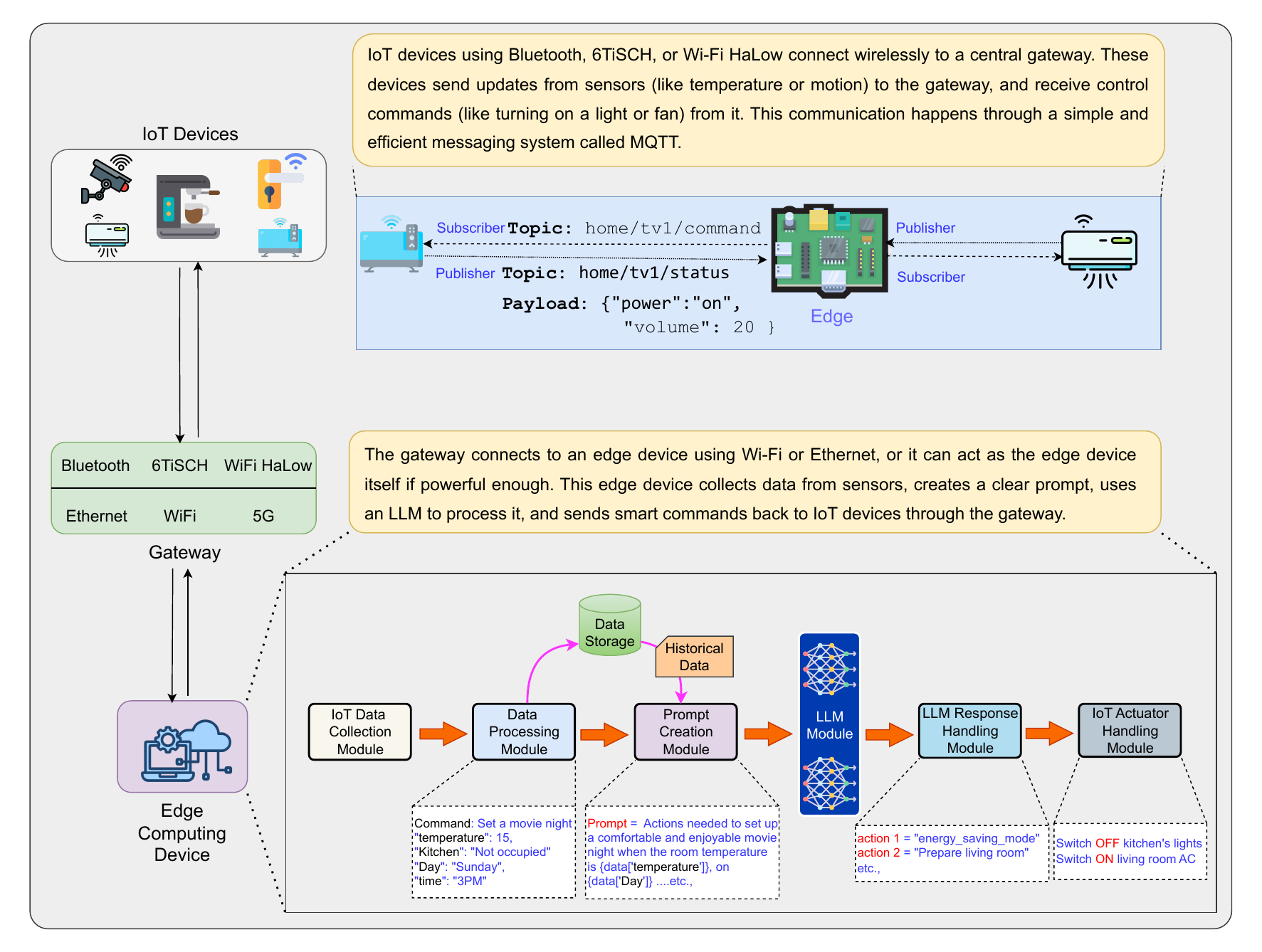}
\caption{Proposed framework is divided into multiple structured modules that handle data collection, processing, prompt creation, response handling, and actuator management.}
\label{fig_1}
\end{figure*}

IoT devices are often resource-constrained regarding processing capacity, memory, and power. Consequently, running LLMs directly on these devices is not technically feasible. To address this challenge, we propose a framework where an IoT network's border router (BR) or gateway is connected to an Edge computing device equipped with sufficient resources to infer user requests using LLMs. 
Therefore, in our proposed framework, LLMs are deployed at the edge of the network, and the edge device is connected to the gateway or BR of the IoT network. IoT devices attached to sensors collect information from the environment and transmit their data to the gateway using multi-hop (mesh topology) or single-hop (star topology) communication networks. The gateway then forwards the data to the edge computing device for processing by the LLM deployed at the edge. Note that some edge devices can also serve as IoT gateways. Apart from the input provided by the IoT devices, users can also provide their input directly to the LLMs in the form of text, video, or voice commands. For example, in smart home environments, users can send voice commands to perform different activities. The transmission of data from the IoT devices to the gateway or edge follows traditional source and channel encoding. 

To simplify our proposed LLM in IoT network framework to facilitate efficient and intelligent interactions within IoT ecosystems, we divide it into multiple structured modules that handle data collection, processing, prompt creation, response handling, and actuator management. These modules are shown in Figure~\ref{fig_1} and discussed as follows,

\textbf{IoT Data Collection Module}: This module is responsible for gathering data from various IoT devices, including the commands received from the users. This module ensures real-time data collection from IoT devices and users. For this, light-weight application layer protocol MQTT (Message Queuing Telemetry Transport) can be used. For example, in a smart home environment, both the Smart TV and the edge device act as publishers and subscribers. The Smart TV publishes its status updates (e.g., power state, volume) to a topic like \texttt{home/tv1/status}, which the edge device subscribes to. Conversely, the edge device sends control commands (e.g., \texttt{turn on/off}) by publishing to a topic such as \texttt{home/tv1/command}, which the Smart TV subscribes to. This bidirectional communication allows seamless monitoring and control using the lightweight MQTT protocol. The broker for the MQTT protocol can be deployed in the edge or gateway of the IoT network.

\textbf{Data Processing Module}: Once the data is collected, it is processed to extract relevant information and convert it into a suitable format for further analysis. This module applies filtering, aggregation, and transformation techniques to ensure the data is ready for the next stages.

\textbf{Storage Module}: This module is responsible for maintaining historical data, which can be fed to the LLM for referencing past information, providing context, and improving the experience of the end users. In brief, in addition to sensory input and user input, the LLM model is fed with historical data for better inference. For instance, when a user issues a voice command to set up a movie night, our proposed framework can consider previous historical data from the user to adjust the temperature. 

\textbf{Strictured Prompt Creation Module}: To enable context-aware reasoning, we incorporate a lightweight \textbf{Retrieval-Augmented Generation} (RAG) mechanism within our system. This module dynamically constructs prompts by retrieving relevant information from two sources: (i) real-time sensor data collected from the connected IoT devices, and (ii) historical context and domain-specific knowledge stored in the Storage Module. By combining live environmental inputs with past observations or rules, the system ensures that the LLM operates with an up-to-date and situationally grounded context. These structured prompts allow the LLM to generate responses that are not only linguistically coherent but also semantically aligned with the current state of the IoT environment. As a result, the LLM’s decisions become more accurate, goal-oriented, and responsive to real-world conditions.

\begin{lstlisting}[language=Python, caption=Structured Prompt Generation]
structured_prompt = (
    f"{session['user_command']}.\n"
    f"Only consider these devices: {', '.join(session['devices'])}.\n"
    f"Current sensor readings: {json.dumps(session['current_sensor_values'])}\n"
    "First, give a 20-word description. Then respond ONLY in the following JSON format:\n"
    "{\n"
    "  \"description\": \"<short description>\",\n"
    "  \"commands\": [\n"
    "    {\"device\": \"<device>\", \"action\": \"<action>\", \"mode\": \"<mode> (optional)\" }\n"
    "  ]\n"
    "}"
)
\end{lstlisting}

\textbf{LLM Module}: The core component of the framework, the LLM Module, utilizes advanced language models to interpret prompts and generate coherent, context-aware responses. This module leverages the capabilities of LLMs to understand and process natural language queries effectively. Some of the open-source LLMs are mentioned in Table~\ref{tab:llm_pi_comparison}.
\begin{lstlisting}[language=Python, caption=An example of using an open-source LLM]
def get_llama_response(prompt):
    url = "http://localhost:11434/api/generate"
    payload = {
        "model": "llama3",
        "prompt": prompt,
        "stream": False
    }
\end{lstlisting}
\begin{table}[htbp]
\centering
\caption{Comparison of Open-Source LLMs}
\label{tab:llm_pi_comparison}
\begin{tabular}{|l|c|c|c|}
\hline
\textbf{Model} & \textbf{Params} & \textbf{Q4 Size} & \textbf{Min RAM} \\
\hline
TinyLlama-1.1B & 1.1B & 0.55--0.7 GB & 1.5 GB \\
StableLM-Zephyr-3B & 3B & 1.8--2.2 GB & 4 GB \\
Phi-2 & 2.7B & 1.5--1.8 GB & 3 GB \\
Gemma-2B & 2B & 1.3--1.6 GB & 3 GB \\
LLaMA-2-7B & 7B & 3.8--4.5 GB & 6 GB \\
Mistral-7B & 7B & 4.0--4.5 GB & 6 GB \\
GPT-J-6B & 6B & 3.0--3.5 GB & 5 GB \\
DistilGPT-2 & 82M & 0.3--0.5 GB & 1 GB \\
\hline
\end{tabular}
\end{table}
\textbf{LLM Response Handling Module}: After generating the structured prompt, it is sent to the deployed LLM, such as LLaMA, for inference. The LLM processes the prompt and returns a textual response that contains the intended actions to be taken on the IoT devices. This raw response is then parsed to extract actionable commands in a predefined \texttt{JSON} like format. The parsing step is essential to ensure the output is structured correctly and interpretable by downstream modules.

\textbf{IoT Actuator Handling Module}: The final module in the framework, the IoT Actuator Handling Module, translates the processed responses into actionable commands for IoT devices. It ensures that the system's outputs are effectively communicated within the IoT network, enabling intelligent and automated control.
\begin{lstlisting}[language=Python, caption=Transmitting commands to the IoT devices using MQTT]
def control_device(device, action):
    if device == "fan":
        publish_fan(action)
    elif device == "light":
     ...
\end{lstlisting}

Note that these modules can be easily integrated to work with current IoT infrastructures, enhancing their capabilities without designing the entire IoT system from scratch. For example, an LLM can interface with existing smart home assistants and IoT devices, using APIs and standard communication protocols to understand and execute user commands.

\section{Experimental Setup and Evaluation}

To validate the feasibility of deploying LLMs at the edge of IoT networks for natural-language-based control, we developed a smart home prototype using a Raspberry Pi 5 (8 GB) as the edge computing device. Three appliances \textit{i.e.,} a light, a TV and a fan were connected to the edge via \texttt{ESP8266 NodeMCU} microcontrollers using 5V mechanical \texttt{relay} modules. Communication between the Raspberry Pi and the IoT nodes was established over Wi-Fi using the MQTT protocol.

Each IoT device is subscribed to a unique MQTT topic for receiving control commands and publishes its current status to a dedicated topic. The user issued natural language commands through a local text-based interface, which were converted into structured prompts and processed by the LLMs. The resulting response was parsed into \texttt{JSON} format and translated into MQTT commands, which were then sent to the appropriate devices. We tested two different LLMs i.e., \textbf{LLaMA 3 (7B)} and  \textbf{Gemma 2B}. Each model was evaluated with three representative commands related to smart room setup. For each command, we recorded the LLM’s response and the total inference time from prompt submission to command dispatch. The expected device states were pre-defined for each use case. Table~\ref{tab:llm-comparison} summarizes the results.
\begin{table*}[t]
\centering
\caption{Comparison of LLaMA 3 and Gemma 2B on Smart Home environment }
\label{tab:llm-comparison}
\resizebox{\textwidth}{!}{%
\begin{tabular}{|p{1.2cm}|p{5cm}|p{4cm}|p{4cm}|c|}
\hline
\textbf{Model} & \textbf{Command} & \textbf{Expected Output} & \textbf{LLM Response} & \textbf{Time (s)} \\
\hline
\multirow{3}{*}{LLaMA 3} 
& 1. \texttt{Set the room for Study}     
    & Light: \texttt{on}, Fan: \texttt{on}, TV: \texttt{off} 
    & Light: \texttt{on}, Fan: \texttt{on}, TV: \texttt{off} 
    & 208 \\
& 2. \texttt{Set the room for movie night} 
    & Light: \texttt{on}, Fan: \texttt{on}, TV: \texttt{on} 
    & Light: \texttt{on}, Fan: \texttt{on}, TV: \texttt{on} 
    & 204 \\
& 3. \texttt{I want to sleep now}        
    & Light: \texttt{off}, Fan: \texttt{off}, TV: \texttt{off} 
    & Light: \texttt{\textcolor{red}{on}}, Fan: \texttt{off}, TV: \texttt{off} 
    & 126 \\
\hline
\multirow{3}{*}{Gemma 2B} 
& 1. \texttt{Set the room for Study}     
    & Light: \texttt{on}, Fan: \texttt{on}, TV: \texttt{off} 
    & Light: \texttt{on}, Fan: \texttt{on}, TV: \texttt{\textcolor{red}{on}} 
    & 28 \\
& 2. \texttt{Set the room for movie night} 
    & Light: \texttt{on}, Fan: \texttt{on}, TV: \texttt{on} 
    & Light: \texttt{on}, Fan: \texttt{on}, TV: \texttt{\textcolor{red}{off}} 
    & 29 \\
& 3. \texttt{I want to sleep now}        
    & Light: \texttt{off}, Fan: \texttt{off}, TV: \texttt{off} 
    & Light: \texttt{off}, Fan: \texttt{off}, TV: \texttt{off} 
    & 30 \\
\hline
\end{tabular}%
}
\end{table*}
LLaMA 3 exhibited high semantic accuracy, generating responses that closely matched the expected device states for each user command. However, its inference time was significantly higher, ranging from \textit{126} to \textit{208 seconds}. Note that these delays define the execution times of the LLMs only. We do not consider the other delays like propagation and transmission delays of the command, delay for generating structured prompts, and so on. In contrast, Gemma 2B demonstrated significantly lower latency (\textit{28–30 seconds}) across all commands. However, it occasionally misinterpreted the intent, such as turning on the TV when it should remain off or failing to activate it when required. This suggests that while lightweight models are more suitable for low-latency edge deployment, they may require fine-tuning or optimized prompt engineering to achieve comparable accuracy to larger models. Overall, the results indicate that LLMs can be effectively integrated into edge-based IoT control systems. LLaMA 3 ensures high accuracy but requires more compute and response time, whereas Gemma 2B offers faster, local inferencing at the cost of occasional semantic deviations. This trade-off between inference quality and latency is crucial when choosing models for real-time IoT applications. The output of one user command ``\texttt{I want to sleep now}'' is shown below.

\begin{lstlisting}[language=Python]
# User Input
Enter your smart home command: I want to sleep now

# Sending to LLaMA...
Raw Output from LLaMA:
Here is your sleep-aiding response:
Description: Prepare your sleep sanctuary with a dim light and calming TV settings.
{
  "description": "Prepare your sleep sanctuary with a dim light and calming TV settings.",
  "commands": [
    {"device": "light", "action": "dim"},
    {"device": "tv", "action": "off"}
    {"device": "fan", "action": "off"}
  ]
}
# Post processing of LLM output
Description: Prepare your sleep sanctuary with a dim light and calming TV settings.

Commands:

Device: light | Desired: dim | Current: on
No action needed for light

Device: tv | Desired: off | Current: on
Turning OFF tv

Device: fan | Desired: off | Current: off
No action needed for fan

# Final Action
TV = Turn Off
\end{lstlisting}

\section{Advantages of Using LLMs in IoT Networks}

\textbf{Improved Decision Making}:  
LLMs enhance decision-making in IoT systems by understanding user intent and environmental context. This allows the system to infer appropriate actions even from vague or high-level inputs. For example, a command like ``\texttt{prepare the room for sleep}'' can lead to multiple coordinated actions such as turning off lights and reducing fan speed, all without needing the user to specify each task individually.

\textbf{Faster Response Time}:  
By processing natural language commands directly at the edge or via lightweight prompt-based inference, LLMs reduce the need for large data exchanges and multi-step rule evaluation. This streamlined processing pipeline reduces transmission, computation, and queuing delays, leading to quicker system responses, especially in time-sensitive IoT applications.

\textbf{Enhanced User Experience}:  
Natural language interfaces powered by LLMs allow users to interact with IoT systems more intuitively and efficiently. Users can issue simple voice or text commands, which are interpreted accurately by the system. By incorporating historical user data, the system can also learn preferences over time, enabling personalized automation and reducing repetitive configurations.

\textbf{Scalability and Adaptability}:  
LLMs provide a flexible control interface that scales with the number of IoT devices and services without requiring complex rule-based programming. As new devices are added to the system, the LLM can dynamically adapt to new vocabulary and control patterns, simplifying deployment in large-scale or evolving environments.

\section{Use Cases}
This section discusses some of the use cases of how LLMs can be integrated into Edge-based IoT systems:

\textbf{Industrial IoT}: In next-generation Industrial IoT (IIoT) environments, including smart manufacturing systems, intelligent energy grids, and autonomous industrial infrastructures, LLMs are expected to play a pivotal role in enhancing system intelligence, adaptability, and autonomy. Future IIoT architectures may employ hierarchical LLM-based frameworks wherein edge-level agents perform real-time sensor fusion and anomaly detection, while cloud-based LLMs synthesize contextual information from historical records, sensor data, and operational knowledge to generate high-level decisions for predictive maintenance and system optimization. These models can be used to enable data-driven forecasting of equipment failures by identifying subtle correlations across multimodal sensor streams, thereby supporting condition-based maintenance and minimizing unplanned downtime. LLMs can also serve as decision-support agents, analyzing continuous operational data such as throughput, energy consumption, and fault indicators to dynamically recommend adjustments that improve process efficiency, safety, and resource allocation. Moreover, LLMs are expected to facilitate natural language interfaces for industrial systems, allowing operators to issue queries or commands in plain language, which the model can translate into machine-executable actions. This capability addresses long-standing challenges in human-machine interaction by improving accessibility and reducing interface complexity. With advancements in multimodal reasoning, future LLMs are envisioned to integrate heterogeneous data sources including time-series signals, visual inputs, textual logs, and expert annotations into a unified semantic framework, thereby enhancing situational awareness and decision accuracy. 
The development of domain-specialized, resource-efficient LLMs tailored for IIoT use cases is expected to significantly improve the robustness, transparency, and operational intelligence of industrial systems deployed across resource-constrained and mission-critical environments.

\textbf{Smart Healthcare}:
In future Internet of Medical Things (IoMT) ecosystems comprising of wearable sensors, remote monitoring devices, smart diagnostics, and hospital infrastructure, LLMs can be expected to enable intelligent, patient-centric healthcare delivery through real-time monitoring, personalized support, and clinical decision assistance. LLMs deployed at the edge could continuously analyze different physiological data (e.g., heart rate, oxygen saturation, glucose levels) to detect early indicators of medical deterioration and generate context-aware alerts, while accounting for patient-reported symptoms and behavioral context. Such systems may leverage LLMs for improving the sensitivity and specificity of automated monitoring. Additionally, conversational LLM-based health assistants can be developed to facilitate natural interaction with patients and caregivers, enabling voice-based symptom reporting, personalized medication reminders, and dynamic health education ultimately improving adherence and reducing clinical workload. 
Furthermore, by integrating health records, lifestyle data, and sensor inputs, LLMs can support predictive analytics and treatment personalization such as identifying high-risk patterns in diabetic patients or forecasting readmission risks in post-operative cases. To mitigate clinician burden and information overload, LLMs can also assist in data summarization, transforming continuous IoMT streams into structured reports that highlight actionable insights. 

\textbf{Semantic Communication}: With the increasing deployment of IoT devices and the emergence of data-intensive applications, traditional communication systems are facing significant challenges related to bandwidth efficiency, latency, and scalability. As current physical-layer technologies approach Shannon’s limit, semantic communication, which focuses on transmitting the meaning rather than the raw data has emerged as a promising paradigm \cite{10328186}. By integrating LLMs into IoT systems, particularly at the network edge, intelligent, context-aware, and bandwidth-efficient communication can be enabled. For example, the door security cameras in smart homes, instead of transmitting a full high-resolution image to a central server for processing, the edge device can extract semantic features such as image edges and natural language descriptions like ``\textit{a fair-colored \textbf{unknown} man wearing a blue t-shirt with a tense face}''. These compact representations consume significantly less bandwidth and processing power. Upon reception, the LLM at the receiver end can reconstruct or interpret the image using the edge-map and description, potentially even regenerating a high-fidelity approximation of the original scene using generative models. This approach drastically reduces transmission load, minimizes network congestion, and lowers end-to-end latency \cite{kalita2024largelanguagemodelsllms}. Beyond compression, LLMs enable context-aware summarization, anomaly detection, and event-triggered communication, ensuring that only meaningful and prioritized information is transmitted. When deployed at the edge, such systems can adapt in real-time, leveraging historical and situational data to improve decision-making accuracy and optimize communication schedules. Therefore, in the context of 5G and emerging 6G technologies, it is expected that LLM-powered semantic communication holds the potential to revolutionize IoT networking by shifting the focus from syntactic data accuracy to semantic relevance, thereby enhancing the efficiency, scalability, and responsiveness of future smart systems.

\section{Open Challenges and Issues}
Even though LLMs in Edge-based IoT systems can revolutionize the current IoT system, there are a variety of challenges to be tackled before they can be used in real applications. This section discusses the two major open issues for future investigations.

\subsection{Privacy and Security}

The integration of LLMs into IoT systems introduces critical concerns on privacy. Such systems frequently collect highly sensitive data ranging from in-home camera footage to personal health metrics and daily behavioral patterns. This raises the fundamental question: \textit{Where is the data processed, and who has access to it}? The answer is inherently tied to the architectural design of the system.

When an LLM is executed locally on an edge device, data processing remains within the user's trusted boundary, offering a high degree of privacy preservation \cite{khatiwada2025}. Conversely, when raw sensor data is transmitted to a cloud-based LLM or a third-party API, there is a non-trivial risk of data exposure, unauthorized storage, or misuse. Despite these risks, cloud-based approaches offer significant advantages in terms of cost-efficiency and deployment scalability, as they benefit from access to extensive datasets and substantial computational resources. While local open-source LLMs may not match the scale of cloud models, their lightweight nature enables fast, private execution and supports a broad range of practical tasks. Therefore, there are trade offs running the LLMs either on edge (local) device or cloud. However, a promising alternative can be a hybrid architectures, where preliminary processing of the data can be done on the edge device, and only encrypted summaries or non-sensitive features can be transmitted to the cloud for advanced inference. This approach provides a balanced trade-off between data privacy and system performance, mitigating the limitations of both extremes.

Security is the complementary challenge in the same context. An LLM deployed at the edge can be prone to different types of attacks. For example, adversaries might attempt to compromise system integrity through techniques such as \textit{prompt injection}, wherein manipulated inputs could lead the LLM to disclose sensitive information or behave undesirably. In extreme cases, if the model retains fragments of user input across sessions, it may inadvertently act as a leakage channel for private data. Mitigating these risks requires robust input filtering mechanism which can detect suspicious input, and sandboxing mechanisms to restrict the model’s access and operational domain.

In summary, ensuring user trust in LLM-enabled IoT systems necessitates rigorous attention to privacy and security. 
In environments where IoT devices are continuously sensing, recording, and interpreting user data, it is imperative that the intelligence driving them adheres to the same ethical and operational standards expected of any trusted human assistant.

\subsection{Accuracy and Reliability in LLM-Enabled IoT Systems}

In safety-critical domains such as industrial automation, healthcare, and smart infrastructure, the reliability of LLM-generated outputs is essential. Unlike general-purpose chatbot applications, erroneous outputs in IoT settings can have physical consequences such as triggering false alarms, missing critical faults, or causing unsafe actuator behavior. For instance, in a smart home environment, if an LLM misinterprets smoke sensor data as non-threatening, it may delay emergency response to an actual fire hazard.  Conversely, an overly sensitive model might trigger frequent negative actuation, such as turning off appliances unnecessarily or repeatedly notifying medical staff for non-critical fluctuations leading to alert fatigue and reduced trust in the system.

Therefore, ensuring reliable behavior in different environments requires a multi-layered approach. Fine-tuning LLMs on domain-specific data (e.g., annotated sensor logs and fault descriptions) or layer-wise unified compression reduces hallucinations and enhances context-aware accuracy \cite{birkmos,yu2024edge}. In addition, hybrid architectures that pair LLMs with deterministic rule-based verifiers can flag or override unsafe critical decisions. For instance, LLM-generated commands such as shutting down a compressor should be validated against physical thresholds before execution.

Furthermore, to properly evaluate the performance of LLM based IoT systems, we need benchmarks that go beyond just language quality. These benchmarks should also check how accurately the models handle real-world tasks, how fast they respond, and how well they deal with unusual or unexpected situations. In summary, while LLMs bring powerful reasoning capabilities to IoT, their deployment in critical applications must be underpinned by domain adaptation, architectural safeguards, transparent outputs, and rigorous evaluation. With these measures, LLMs can become dependable agents capable of responsible decision-making in real-world IoT systems.

\section{Conclusion}

In this article, we discussed how LLMs can enhance the efficiency and intelligence of edge-based IoT systems. We proposed a modular framework that integrates LLMs into IoT architectures and outlined the role of each component within the system, from data collection and prompt generation to response handling and device control. We also explored the advantages of using LLMs for natural language-based interaction, personalized automation, and context-aware decision-making in IoT environments. Finally, we highlighted a few use cases of LLM-based IoT systems, key challenges such as privacy, security, reliability, and model optimization that must be addressed to enable safe and effective deployment of LLM-enabled IoT systems in real-world scenarios.

\bibliographystyle{IEEEtran}
\bibliography{refs}

\begin{thebibliography}{10}
\providecommand{\url}[1]{#1}
\csname url@samestyle\endcsname
\providecommand{\newblock}{\relax}
\providecommand{\bibinfo}[2]{#2}
\providecommand{\BIBentrySTDinterwordspacing}{\spaceskip=0pt\relax}
\providecommand{\BIBentryALTinterwordstretchfactor}{4}
\providecommand{\BIBentryALTinterwordspacing}{\spaceskip=\fontdimen2\font plus
\BIBentryALTinterwordstretchfactor\fontdimen3\font minus \fontdimen4\font\relax}
\providecommand{\BIBforeignlanguage}[2]{{%
\expandafter\ifx\csname l@#1\endcsname\relax
\typeout{** WARNING: IEEEtran.bst: No hyphenation pattern has been}%
\typeout{** loaded for the language `#1'. Using the pattern for}%
\typeout{** the default language instead.}%
\else
\language=\csname l@#1\endcsname
\fi
#2}}
\providecommand{\BIBdecl}{\relax}
\BIBdecl

\bibitem{10.1145/3536166}
A.~Kalita and M.~Khatua, ``{6TiSCH – IPv6 Enabled Open Stack IoT Network Formation: A Review},'' \emph{ACM Trans. Internet Things}, vol.~3, no.~3, jul 2022.

\bibitem{kok2024}
\BIBentryALTinterwordspacing
I.~Kok, O.~Demirci, and S.~Ozdemir, ``{When IoT Meet LLMs: Applications and Challenges},'' 2024. [Online]. Available: \url{https://arxiv.org/abs/2411.17722}
\BIBentrySTDinterwordspacing

\bibitem{gao2024}
\BIBentryALTinterwordspacing
Y.~Gao, Z.~Ye, M.~Xiao, Y.~Xiao, and D.~I. Kim, ``{Guiding IoT-Based Healthcare Alert Systems with Large Language Models},'' 2024. [Online]. Available: \url{https://arxiv.org/abs/2408.13071}
\BIBentrySTDinterwordspacing

\bibitem{lin2024pushinglargelanguagemodels}
\BIBentryALTinterwordspacing
Z.~Lin, G.~Qu, Q.~Chen, X.~Chen, Z.~Chen, and K.~Huang, ``{Pushing Large Language Models to the 6G Edge: Vision, Challenges, and Opportunities},'' 2024. [Online]. Available: \url{https://arxiv.org/abs/2309.16739}
\BIBentrySTDinterwordspacing

\bibitem{an2025}
\BIBentryALTinterwordspacing
T.~An, Y.~Zhou, H.~Zou, and J.~Yang, ``{IoT-LLM: Enhancing Real-World IoT Task Reasoning with Large Language Models},'' 2025. [Online]. Available: \url{https://arxiv.org/abs/2410.02429}
\BIBentrySTDinterwordspacing

\bibitem{10342693}
A.~Hazra, A.~Kalita, and M.~Gurusamy, ``{Meeting the Requirements of Internet of Things: The Promise of Edge Computing},'' \emph{IEEE Internet of Things Journal}, vol.~11, no.~5, pp. 7474--7498, 2024.

\bibitem{10328186}
S.~Barbarossa, D.~Comminiello, E.~Grassucci, F.~Pezone, S.~Sardellitti, and P.~Di~Lorenzo, ``{Semantic Communications Based on Adaptive Generative Models and Information Bottleneck},'' \emph{IEEE Communications Magazine}, vol.~61, no.~11, pp. 36--41, 2023.

\bibitem{kalita2024largelanguagemodelsllms}
\BIBentryALTinterwordspacing
A.~Kalita, ``{Large Language Models (LLMs) for Semantic Communication in Edge-based IoT Networks},'' 2024. [Online]. Available: \url{https://arxiv.org/abs/2407.20970}
\BIBentrySTDinterwordspacing

\bibitem{khatiwada2025}
\BIBentryALTinterwordspacing
K.~Khatiwada, J.~Hopper, J.~Cheatham, A.~Joshi, and S.~Baidya, ``{Large Language Models in the IoT Ecosystem -- A Survey on Security Challenges and Applications},'' 2025. [Online]. Available: \url{https://arxiv.org/abs/2505.17586}
\BIBentrySTDinterwordspacing

\bibitem{birkmos}
\BIBentryALTinterwordspacing
R.~Birkmose, N.~M. Reece, E.~H. Norvin, J.~Bjerva, and M.~Zhang, ``{On-Device LLMs for Home Assistant: Dual Role in Intent Detection and Response Generation},'' 2025. [Online]. Available: \url{https://arxiv.org/abs/2502.12923}
\BIBentrySTDinterwordspacing

\bibitem{yu2024edge}
\BIBentryALTinterwordspacing
Z.~Yu, Z.~Wang, Y.~Li, H.~You, R.~Gao, X.~Zhou, S.~R. Bommu, Y.~K. Zhao, and Y.~C. Lin, ``{EDGE-LLM: Enabling Efficient Large Language Model Adaptation on Edge Devices via Layerwise Unified Compression and Adaptive Layer Tuning and Voting},'' 2024. [Online]. Available: \url{https://arxiv.org/abs/2406.15758}
\BIBentrySTDinterwordspacing

\end{thebibliography}

\end{document}